\documentstyle[epsf]{mn}

\newif\ifAMStwofonts

\def\lesssim{\mathrel{\hbox{\rlap{\hbox{\lower4pt\hbox{$\sim$}}}\hbox{$<$}}}}
\def\gtrsim{\mathrel{\hbox{\rlap{\hbox{\lower4pt\hbox{$\sim$}}}\hbox{$>$}}}}
\font\gkvec=cmmib10                         
\def\bgamma{\hbox{{\gkvec\char13}}}        

\def\eps@scaling{.95}
\def\epsscale#1{\gdef\eps@scaling{#1}}
\def\plotone#1{\centering \leavevmode
\epsfxsize=\eps@scaling\columnwidth \epsfbox{#1}}

\ifoldfss
  \ifCUPmtlplainloaded \else
    \NewTextAlphabet{textbfit} {cmbxti10} {}
    \NewTextAlphabet{textbfss} {cmssbx10} {}
    \NewMathAlphabet{mathbfit} {cmbxti10} {} 
    \NewMathAlphabet{mathbfss} {cmssbx10} {} 
  \fi
  \ifAMStwofonts
    \ifCUPmtlplainloaded \else
      \NewSymbolFont{upmath} {eurm10}
      \NewSymbolFont{AMSa} {msam10}
      \NewMathSymbol{\upi}     {0}{upmath}{19}
      \NewMathSymbol{\umu}     {0}{upmath}{16}
      \NewMathSymbol{\upartial}{0}{upmath}{40}
      \NewMathSymbol{\leqslant}{3}{AMSa}{36}
      \NewMathSymbol{\geqslant}{3}{AMSa}{3E}

       \let\le=\leqslant
       \let\ge=\geqslant
    \fi
  \fi
\fi 

\ifnfssone
  \newmathalphabet{\mathit}
  \addtoversion{normal}{\mathit}{cmr}{m}{it}
  \addtoversion{bold}{\mathit}{cmr}{bx}{it}
  \newmathalphabet{\mathbfit} 
  \addtoversion{normal}{\mathbfit}{cmr}{bx}{it}
  \addtoversion{bold}{\mathbfit}{cmr}{bx}{it}
  \newmathalphabet{\mathbfss} 
  \addtoversion{normal}{\mathbfss}{cmss}{bx}{n}
  \addtoversion{bold}{\mathbfss}{cmss}{bx}{n}
  \ifAMStwofonts
    \ifCUPmtlplainloaded \else
      \UseAMStwoboldmath
      \makeatletter
      \new@mathgroup\upmath@group
      \define@mathgroup\mv@normal\upmath@group{eur}{m}{n}
      \define@mathgroup\mv@bold\upmath@group{eur}{b}{n}
      \edef\UPM{\hexnumber\upmath@group}
      \new@mathgroup\amsa@group
      \define@mathgroup\mv@normal\amsa@group{msa}{m}{n}
      \define@mathgroup\mv@bold\amsa@group{msa}{m}{n}
      \edef\AMSa{\hexnumber\amsa@group}
      \makeatother
      \mathchardef\upi="0\UPM19
      \mathchardef\umu="0\UPM16
      \mathchardef\upartial="0\UPM40
      \mathchardef\leqslant="3\AMSa36
      \mathchardef\geqslant="3\AMSa3E

       \let\le=\leqslant
       \let\ge=\geqslant
    \fi
  \fi
\fi 

\ifnfsstwo
  \DeclareMathAlphabet{\mathbfit}{OT1}{cmr}{bx}{it}
  \SetMathAlphabet\mathbfit{bold}{OT1}{cmr}{bx}{it}
  \DeclareMathAlphabet{\mathbfss}{OT1}{cmss}{bx}{n}
  \SetMathAlphabet\mathbfss{bold}{OT1}{cmss}{bx}{n}
  \ifAMStwofonts
    \ifCUPmtlplainloaded \else
      \DeclareSymbolFont{UPM}{U}{eur}{m}{n}
      \SetSymbolFont{UPM}{bold}{U}{eur}{b}{n}
      \DeclareSymbolFont{AMSa}{U}{msa}{m}{n}
      \DeclareMathSymbol{\upi}{0}{UPM}{"19}
      \DeclareMathSymbol{\umu}{0}{UPM}{"16}
      \DeclareMathSymbol{\upartial}{0}{UPM}{"40}
      \DeclareMathSymbol{\leqslant}{3}{AMSa}{"36}
      \DeclareMathSymbol{\geqslant}{3}{AMSa}{"3E}

       \let\le=\leqslant
       \let\ge=\geqslant
    \fi
  \fi
\fi 

\ifCUPmtlplainloaded \else
  \ifAMStwofonts \else 
    \def\upi{\pi}
    \def\umu{\mu}
    \def\upartial{\partial}
  \fi
\fi

\title[Effects of absorption and Faraday rotation on accretion disk
polarization]{ Polarization from magnetized accretion discs:  II. The effects
of absorption opacity on Faraday rotation}
\author[E. Agol et al.]
       {Eric Agol, Omer Blaes, and Cristian Ionescu-Zanetti\\
        Department of Physics, University of California, Santa Barbara, CA 93106}
\date{Accepted .
      Received ;
      in original form }

\pagerange{\pageref{firstpage}--\pageref{lastpage}}
\pubyear{1996}

\begin{document}

\maketitle

\label{firstpage}

\begin{abstract}
Equipartition magnetic fields can dramatically affect the polarization of
radiation emerging from accretion disk atmospheres in active galactic nuclei.
We extend our previous work on this subject by exploring the interaction
between Faraday rotation and absorption opacity in local, plane-parallel
atmospheres with parameters appropriate for accretion discs.  Faraday
rotation in pure scattering atmospheres acts to depolarize the radiation field
by rotating the polarization planes of photons after last scattering.
Absorption opacity in an unmagnetized atmosphere can
increase or decrease the polarization compared to the pure scattering case,
depending on the thermal source function gradient.  Combining both Faraday
rotation and absorption opacity, we find the following results.
If absorption opacity is much larger than scattering opacity throughout
the atmosphere, then Faraday rotation generally has only a small effect
on the emerging polarization because of the small electron column density
along a photon mean free path.  However, if the absorption opacity is not
too large and it acts alone to increase the polarization, then the effects of
Faraday rotation can be enhanced over those in
a pure scattering atmosphere.  Finally, while Faraday rotation often
depolarizes the radiation
field, it can in some cases increase the polarization when
the thermal source function does not rise too steeply with optical depth.
We confirm the correctness of the Silant'ev (1979) analytic calculation
of the high magnetic field limit of the pure scattering atmosphere, which
we incorrectly disputed in our previous paper.

\end{abstract}


\section{Introduction}

Explaining the optical polarization observed in active galactic nuclei (AGN)
has long been a problem for accretion disk models.  Optically thick, pure
electron scattering discs are expected to emit radiation which is linearly
polarized up to 11.7 percent parallel to the plane of the disk (Chandrasekhar
1960), but this has never been observed in type 1 AGN.  One possible reason
is that the optical radiation emerging from the disk is Faraday depolarized
by photospheric magnetic fields.  In a previous paper (Agol \& Blaes 1996,
hereafter paper I), we have shown that such fields can drastically reduce
the polarization at optical wavelengths if they are near equipartition strength.

Absorption opacity can also play a significant role in determining
the polarization of the emerging radiation field.  In a simple investigation
of this effect, Laor, Netzer \& Piran (1990) showed that absorption opacity
can reduce the overall polarization.  A more careful treatment by Blaes \&
Agol (1996) showed that while this is often qualitatively true, absorption
opacity can sometimes increase the overall polarization by increasing the
limb darkening (see also Bochkarev, Karitskaya, \& Sakhibullin 1985).

The Faraday rotation calculations of paper I assumed a pure electron
scattering atmosphere, but absorption opacity might reduce the depolarization.
This is because the Faraday rotation of a given photon depends on the
total electron column density that the photon traverses.
The dominant effect of Faraday rotation occurs after last scattering (paper I),
so if the absorption opacity significantly reduces the electron scattering
column down to unit optical depth, the depolarization would be smaller.
On the other hand, the absorption opacity itself may directly increase or
decrease the polarization from an unmagnetized disk, as noted above.
In this paper we attempt to disentangle these effects in order to understand
how Faraday rotation and absorption opacity act together to determine the
polarization of the radiation emerging from AGN accretion discs.  We have
discovered a number of subtle phenomena which are not immediately obvious from
the above arguments.

As in paper I, we use Monte Carlo calculations of the radiative transfer.
In addition, however, we also show in section 2 below how Faraday rotation
by a uniform, vertical magnetic field can be incorporated directly into
the radiative transfer equation.  The emerging radiation field can then be
calculated much faster using standard finite difference techniques, and we
present the results of both approaches in simple toy atmosphere models in
section 3.  In section 4 we discuss again the role of Faraday rotation in
determining the optical polarization in AGN accretion discs, and we summarize
our conclusions in section 5.

\section{Equations and Numerical Techniques}

To calculate the radiation field emerging from the accretion disk, we
treat each portion of the disk photosphere as a locally plane-parallel,
semi-infinite
atmosphere.  At the optical and ultraviolet photon frequencies of interest,
electron scattering in the magnetized plasma has negligible circular
dichroism, so the polarization of the radiation field must be nearly linear (cf.
paper I).\footnote{
Whitney (1991a, 1991b) has conducted Monte Carlo calculations of the
polarization of a magnetized, electron scattering atmosphere.  Her calculations
provide an interesting contrast to ours, because she included magnetic
corrections to the scattering cross-section, but neglected Faraday rotation.
Her results are of relevance to magnetic white dwarf and neutron star
atmospheres.  Our work has neglected magnetic effects on the scattering
cross-section but has included Faraday rotation.  This is much more relevant
to optical and ultraviolet radiation emerging from AGN accretion discs, because
the corrections to the scattering cross-section are negligible.}~

\subsection{Monte Carlo Calculations}

Let ${\bf p}$ be a unit vector in the plane of polarization
of a given photon, perpendicular to its direction of propagation ${\bf n}$,
also a unit vector.  Between scatterings, the photon polarization will be
Faraday rotated to
\begin{equation}
\label{eqprot}
{\bf p}_{\rm rot}={\bf p}\cos\chi+({\bf n}\times{\bf p})\sin\chi,
\end{equation}
where $\chi\equiv{\bf b}\cdot{\bf n}\tau_T\delta/2$, $\tau_T$ is the Thomson
scattering depth along the photon trajectory, and ${\bf b}\equiv{\bf B}/B$ is
a unit vector along the magnetic field ${\bf B}$.  The photon wavelength
$\lambda$ and magnetic field strength only enter through the parameter
\begin{equation}
\delta\equiv{3B\lambda^2\over8\pi^2e}\simeq 0.198\left({\lambda\over5000
{\rm\AA}}\right)^2\left({B\over1{\rm G}}\right),
\end{equation}
where $e$ is the electron charge.

Paper I describes a Monte Carlo technique based on these equations to
calculate the polarized radiative transfer through a magnetized, pure
electron scattering atmosphere.  We have modified this slightly to include the
effects of absorption opacity $\kappa_\nu$ at frequency $\nu$, assuming for
simplicity that the ratio of absorption opacity to electron scattering opacity
is independent of optical depth in the atmosphere.  In other words,
\begin{equation}
q_\nu\equiv{n_e\sigma_T\over\kappa_\nu+n_e\sigma_T}
\end{equation}
is constant, where $n_e$ is the electron number density and $\sigma_T$ is
the Thomson cross section.

We propagate each photon a vertical optical depth
$\tau_\nu=\tau_{\nu0}+\mu\ln(r_1)$ through the atmosphere, where $r_1$ is a
random deviate between 0 and 1, $\mu$ is the direction cosine of the photon
propagation vector with respect to the upward vertical, and $\tau_{\nu0}$ is
the starting optical depth.  The photon's polarization vector is Faraday
rotated according to equation (\ref{eqprot}).  Then, another random deviate,
$r_2$, between 0 and 1 is chosen.  If $r_2$ is less than $q_\nu$, the
photon is scattered.  Otherwise it is absorbed and another photon is
started at the base of the atmosphere.  This process is repeated
until a photon escapes from the atmosphere, and it is binned as
described in paper I.

\subsection{Feautrier Radiative Transfer}

One of the advantages of the Monte Carlo technique is that it is capable
of handling general, complex geometries.  Because our purpose in this
paper is to understand the physical effects of Faraday rotation in the
presence of absorption opacity, we limit consideration to locally plane-parallel
atmospheres with uniform, vertical magnetic field.  The radiation field will
then be completely axisymmetric and depend only on vertical depth.  In this
case it is straightforward to include the Faraday rotation directly in
the full radiative transfer equation by just adding an extra term.  This
equation can then be solved much more quickly using standard numerical
techniques.

The full polarized radiative transfer equation for
a general magnetoactive plasma is already well-known (see e.g. Silant'ev 1979).
However, because in our case it is so simple and illuminates the physics, we
now briefly sketch a derivation of the Faraday rotation term.

We first project the photon polarization vector on two orthogonal axes which
are perpendicular to the propagation direction ${\bf n}$.  Let the first axis
be parallel to the plane of the atmosphere, and the corresponding polarization
vector component be $p_0$.  Let the polarization vector component with respect
to the second axis be $p_{90}$.  Then equation (\ref{eqprot}) implies that
after Faraday rotation,
\begin{equation}
\label{eqprot0}
p_{0{\rm rot}}=p_0\cos\chi-p_{90}\sin\chi
\end{equation}
and
\begin{equation}
\label{eqprot90}
p_{90{\rm rot}}=p_0\sin\chi+p_{90}\cos\chi.
\end{equation}

Following Chandrasekhar (1960), define $I_{r\nu}$ and $I_{l\nu}$ as the
intensities of the radiation corresponding to $p_0$ and $p_{90}$, respectively.
The Stokes parameter $Q_\nu$ may them be defined as $I_{r\nu}-I_{l\nu}$.  In a
similar fashion, let $U_\nu$ be the Stokes parameter with respect to two axes
rotated by 45 degrees from those defined previously.  The Stokes parameter
$V_\nu$ vanishes because the radiation is linearly polarized.  Expressed in
terms of the total specific intensity $I_\nu=I_{r\nu}+I_{l\nu}$ and averages
over the individual polarization vectors of the corresponding photons, we have
$Q_{\nu}=I_\nu(<p_0^2>-<p_{90}^2>)$ and $U_\nu=I_\nu<2p_0p_{90}>$.
The degree of polarization is
\begin{equation}
P={(Q_\nu^2+U_\nu^2)^{1/2}\over I_\nu}.
\end{equation}
(Note that, in contrast to paper I, we have not normalized the Stokes
parameters by the total intensity.)

We now describe the intensity and polarization of the radiation field with
the column vector
\begin{equation}
{\bf I}_\nu(\tau_\nu,\mu)=\pmatrix{I_{\nu}\cr Q_{\nu}\cr U_\nu\cr}.
\end{equation}
From equations (\ref{eqprot0}) and (\ref{eqprot90}), Faraday rotation 
transforms the radiation field according to
\begin{equation}
\label{eqrotmat}
{\bf I}_{\nu {\rm rot}}=\pmatrix{1 & 0 & 0\cr
0 & \cos2\chi & -\sin2\chi\cr 0 & \sin2\chi &
\cos2\chi\cr}{\bf I}_\nu.
\end{equation}
Let $z$ be the height measured vertically upward in the atmosphere.  Then
for an infinitesimal change in height $dz$, the corresponding Faraday rotation
angle for a vertical magnetic field is
\begin{equation}
\label{diffrotang}
d\chi=\mp{1\over2}\delta n_e\sigma_Tdz,
\end{equation}
where the upper (lower) sign is to be taken for an upward
(downward) directed field.  Expanding equation (\ref{eqrotmat}), we deduce
that the effect of Faraday rotation by a vertical magnetic field can be
described by
\begin{equation}
{\partial{\bf I}_\nu\over\partial z}=n_e\sigma_T{\bf F}{\bf I}_\nu,
\end{equation}
where
\begin{equation}
{\bf F}\equiv\pm\delta\pmatrix{0&0&0\cr 0&0&1\cr 0&-1&0\cr}.
\end{equation}
Note that ${\bf F}$ does not change the total intensity, as expected.

Inserting this term into the full radiative transfer equation, we have
\begin{eqnarray}
\mu{\partial{\bf I}_\nu\over\partial z}&=&\eta_\nu\pmatrix{1\cr 0\cr 0\cr}
-(\kappa_\nu+n_e\sigma_T){\bf I}_\nu+\mu n_e\sigma_T{\bf F}{\bf I}_\nu
\nonumber \\
& &+{3\over8}n_e\sigma_T\int_{-1}^1d\mu'{\bf P}(\mu,\mu'){\bf I}_\nu(\mu'),
\end{eqnarray}
where $\eta_\nu$ is the thermal emission coefficient (assumed unpolarized),
\begin{eqnarray}
\lefteqn{{\bf P}(\mu,\mu')\equiv}\nonumber\\
\lefteqn{\pmatrix{{4\over3}\left[1+{1\over2}P_2(\mu)P_2(\mu')
\right] & (1-\mu'^2)P_2(\mu) & 0\cr (1-\mu^2)P_2(\mu') & {3\over2}(1-\mu^2)
(1-\mu'^2) & 0\cr 0 & 0 & 0\cr},}
\end{eqnarray}
and $P_2(\mu)\equiv(3\mu^2-1)/2$ is a second order Legendre polynomial
(Chandrasekhar 1960, Loskutov \& Sobolev 1979).  Switching to the
total optical depth $\tau_\nu$ as the dependent variable in the usual way,
\begin{eqnarray}
\label{eqtransfer}
\mu{\partial{\bf I}_\nu\over\partial\tau_\nu}&=&{\bf I}_\nu-S_\nu(1-q_\nu)
\pmatrix{1\cr 0\cr 0\cr}
-\mu q_\nu{\bf F}{\bf I}_\nu
\nonumber \\
& &-{3\over8}q_\nu\int_{-1}^1d\mu'{\bf P}(\mu,\mu'){\bf I}_\nu(\mu'),
\end{eqnarray}
where $S_\nu\equiv\eta_\nu/\kappa_\nu$ is the thermal source function.

The formal solution for equation (\ref{eqtransfer}) can be expressed in 
terms of the total source function:
\begin{equation}
\bmath{\Im}\equiv\pmatrix{\Im_I \cr \Im_Q \cr \Im_U}\equiv
S(1-q)\pmatrix{1 \cr 0 \cr 0}+
{3\over8}q\int_{-1}^1d\mu'{\bf P}(\mu,\mu'){\bf I}(\mu').
\end{equation}
(Note that $\Im_U=0$ and $\Im_Q$ only has a contribution from scattering.)
Then, the formal solution is given by:
\begin{equation}
\label{eqformfar}
{\bf I}(0,\mu)={\int_0^\infty\pmatrix{1 & 0 & 0 \cr 0 & \cos{t\delta q}
& \pm\sin{t\delta q} \cr 0 & \mp\sin{t\delta q} & \cos{t\delta q}}
\bmath{\Im}(t,\mu)e^{-t/\mu}{dt\over\mu}},
\end{equation}
where the sign convention is the same as for equation (\ref{diffrotang}).
The matrix represents the effect of Faraday rotation from the point of emission
or last scattering to the top of the atmosphere (cf. eq. {\ref{eqrotmat}}).

We have applied the Feautrier technique (e.g. Mihalas \& Mihalas 1984,
Phillips \& M\'esz\'aros 1986) to solving equation
(\ref{eqtransfer}) subject to the boundary condition that there be no external
illumination of the atmosphere at $\tau_\nu=0$.  Unless otherwise noted,
we calculate the integrals
over $\mu$ with sixteen point Gaussian quadratures and use a logarithmically
spaced grid in $\tau$.  

\section{Polarization from Constant \lowercase{$q_\nu$} Atmospheres}

In order to illuminate the physics, we now consider two idealized atmosphere
problems, both with $q_\nu$ independent of optical depth.  The first case has
zero thermal source function everywhere except for a source at infinite
optical depth.  The second case has an isotropic thermal source function which
varies linearly with optical depth.  These problems were solved in the
unmagnetized case by Loskutov \& Sobolev (1979).  Photons of different
frequency are completely decoupled in our radiative transfer equation for an
atmosphere of fixed assumed structure, i.e. there is no frequency
redistribution.  We therefore drop the subscript $\nu$ on all
variables and parameters from now on.

\subsection{Case 1:  Radiation Sources from Infinite Optical Depth}

This problem is a generalization of the pure electron scattering case considered
by Chandrasekhar (1960), but it is important to note that the presence of
absorption opacity in this case implies that the intensity of the radiation
emerging from the top of the atmosphere is formally zero unless there is
infinite illumination from below.  Nevertheless the
radiation field is still polarized.  This problem therefore represents an
idealization of an atmosphere in which photons of a given frequency are
thermally emitted in significant quantities only at large optical depth.

From the numerical standpoint we have performed both the Monte Carlo
simulations
and the Feautrier calculation using sources at sufficiently high, but finite,
optical depth so that the polarization no longer depends on this depth.  
We apply a lower boundary condition of unpolarized, isotropic radiation 
sources. The results then depend on only two parameters, $\delta$ and $q$.

Figure 1 shows the polarization as a function of viewing angle for a variety
of values of $q$ in an unmagnetized atmosphere.  
\begin{figure}
\plotone{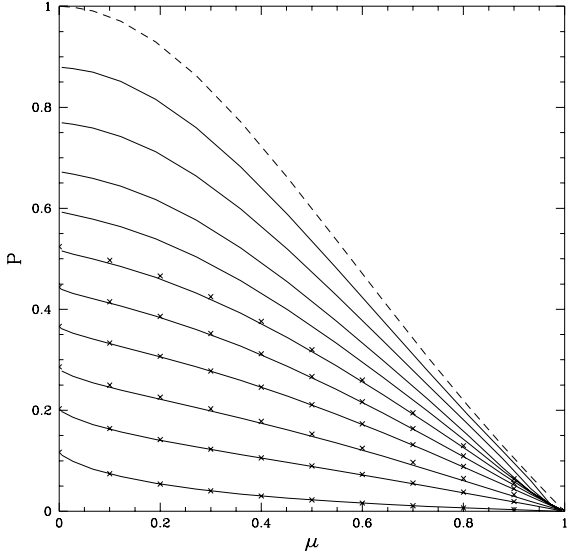}
\caption{Polarization as a function of viewing angle for an unmagnetized
($\delta=0$) atmosphere with different values of $q$ and all radiation sources
at infinite optical depth.  From top to bottom,
the curves represent the results of our Feautrier calculations for $q=0.1$
to $q=1$ in steps of 0.1.  Points represent
the numerical results of Loskutov \& Sobolev (1979), which are consistent
with our results.  The dashed line represents
our analytic solution for $q\rightarrow0$.}
\end{figure}
The lowest curve shows the
pure electron scattering case ($q=1$).  Loskutov \& Sobolev (1979) numerically
calculated cases for $q>0.5$, and found that the polarization increased
monotonically as $q$ decreased.  Their results are also shown in figure 1
and are in excellent agreement with ours.  They also found that the polarization
should continue to rise for even smaller $q$, and we again confirm this fact.
Physically this is somewhat puzzling, because it suggests that the polarization
remains finite even in the $q\rightarrow0$ limit where there is no scattering
opacity.

We have obtained the following $q\rightarrow0$ analytic solution to this
problem in the unmagnetized case.  For outward rays ($0\le\mu\le1$), 
\begin{equation}
\label{eqlosa}
I(\tau,\mu)={(1+\mu^2)e^\tau\over(1-\mu)e^{4/3q}+2}I(0,1),
\end{equation}
and
\begin{equation}
Q(\tau,\mu)=(1+\mu)e^{\tau-4/3q}I(0,1).
\end{equation}
For inward rays ($-1\le\mu<0$),
\begin{equation} 
I(\tau,\mu)={(1+\mu^2)(e^\tau-e^{\tau/\mu})\over(1-\mu)}e^{-4/3q}I(0,1),
\end{equation}
and 
\begin{equation} 
Q(\tau,\mu)=(1+\mu)(e^{\tau}-e^{\tau/\mu})e^{-4/3q}I(0,1).
\end{equation}
The Stokes parameter $U$ vanishes because $\delta=0$.  This
solution may be verified directly by substitution in equation
(\ref{eqformfar}).
In this limit the emergent intensity vanishes except in the upward vertical
direction
($\mu=1$), i.e. there is absolute limb darkening.  The polarized flux
which is represented by $Q$ vanishes for all viewing angles, consistent with
the fact that there is no scattering opacity.  The degree of
polarization does not vanish, however, except along the vertical ($\mu=1$)
because of symmetry:
\begin{equation}
\label{eqlosb}
P={Q\over I}={1-\mu^2\over1+\mu^2}.
\end{equation}
This polarization is also plotted in figure 1 as the dashed line.

For magnetized atmospheres with finite $\delta$, equations
(\ref{eqlosa})-(\ref{eqlosb}) still represent the solution for the radiation
field in the $q\rightarrow0$ limit.  This is because the Faraday rotation term
in equation (\ref{eqtransfer}) is proportional to $q$.  In other words, Faraday
rotation depends on the electron density, and therefore must have negligible
effect when absorption dominates over electron scattering.  This is true
even though this electron scattering produces a nonzero degree of polarization.

Figure 2 shows the polarization for $q=0.8$ and $q=0.2$ atmospheres with
various magnetic field strengths represented by $\delta$.  
\begin{figure}
\plotone{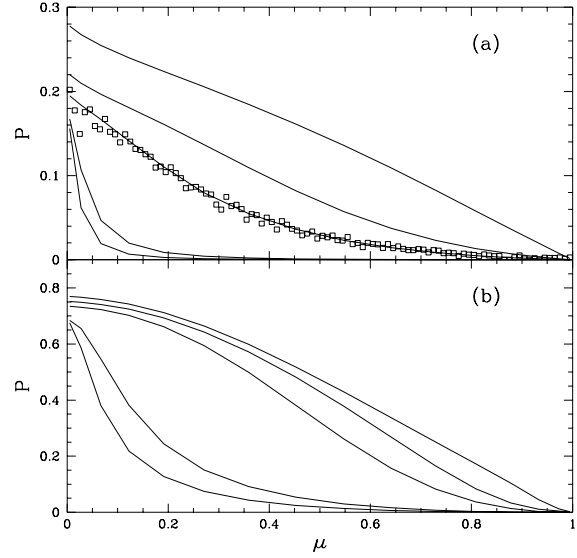}
\caption{Polarization as a function of viewing angle for (a) $q=0.8$ and
(b) $q=0.2$
atmospheres with various values of $\delta$ and all radiation sources at
infinite optical depth.  From top to bottom,
the curves represent the results of our Feautrier calculations for $\delta=0$
(i.e. zero magnetic field), 2, 5, 50, and 100.  Square points represent the
results of our Monte Carlo calculations for the case $q=0.8$ and $\delta=5$,
and confirm the Feautrier results.
}
\end{figure}
As expected, Faraday rotation depolarizes the radiation field.  This figure
should be compared to figure 2(b) of paper I which shows the same thing for a
$q=1$ (zero absorption, pure electron scattering) atmosphere.  It is apparent
that moderate absorption opacity (e.g. $q=0.8$) {\it enhances} the
depolarizing effects of the magnetic field, even though the electron
scattering depth down to unit optical depth is smaller.  This is true
even along lines of sight which are perpendicular to the magnetic field
($\mu=0$).

Based on our Monte Carlo results in paper I, we had claimed that Faraday
rotation by a vertical magnetic field appeared to have no effect on the
polarization for $\mu=0$, at least for the $q=1$ case considered there.  We
rationalized this result on physical grounds by noting that Faraday rotation
probably had its primary effect on photons after last scattering, and such
photons would not suffer any rotation if they travel perpendicular to the
magnetic field.  However, we have repeated the $q=1$ calculations with our
Feautrier code and have found that the polarization at $\mu=0$ drops
from 11.7 per cent at $\delta=0$ (the value from Chandrasekhar 1960) to 9.14
per cent
as $\delta\rightarrow\infty$.  This agrees with the analytic calculation
of Silant'ev (1979), which we wrongly disputed in paper I.  Our Monte Carlo
simulations did not have the resolution to see this decrease since the number
of photons at $\mu=0$ is so small.

Figure 3 shows a comparison of our numerical results with the high $\delta$
calculation of Silant'ev (1979).
\begin{figure}
\plotone{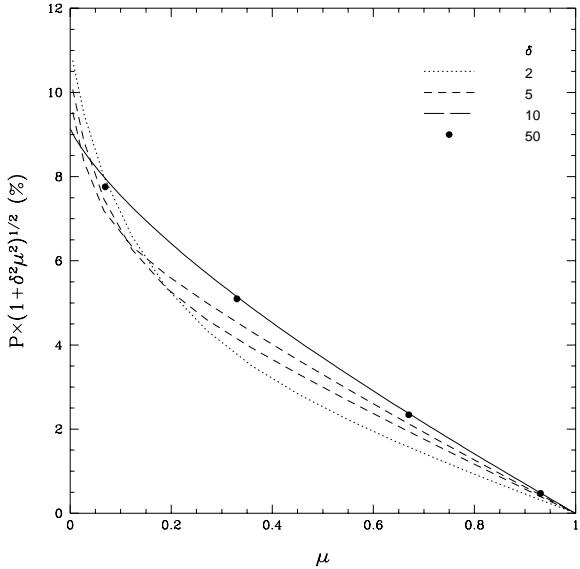}
\caption{Polarization multiplied by $(1+\mu^2\delta^2)^{1/2}$ for various
$\delta$'s.  The solid curve represents Silant'ev's formula for large $\delta$,
while the dashed curves and points depict the results of our Feautrier code.}
\end{figure}
It is quite challenging to calculate the high $\delta$ case numerically.
For $\delta=50$, we reached convergence only with 20,000 logarithmically
spaced depth points from $\tau=10^{-5}$ to $\tau=10$ for 8 angular points.
Even larger $\delta$ becomes numerically prohibitive
because of the large number of matrices that need to be stored 
in the Feautrier method.  In any case we do find agreement with Silant'ev's
formula for large $\delta$.  (In the appendix we show how Silant'ev's result
can be derived from our radiative transfer equation {\ref{eqtransfer}}.)

Since our Feautrier calculations are more accurate than the $q=1$ Monte Carlo
calculations from paper I, we can assess the accuracy of the analytic
fitting formulae found in paper I.  For the $q=1$ case with $\delta\le10$, our
fitting formulae are accurate to better than 12 per cent for $P$, 9 per cent 
for $Q$, and
35 per cent for $U$ (it is least accurate when is very small).  Silant'ev's 
formula 
can be used when $\delta\ge10$, where it is accurate to better than 
11 per cent for $P$.

Faraday rotation generally has only a very small effect on the limb darkening
of the total intensity of the radiation field emerging from an atmosphere,
as shown in figure 4.  In the pure scattering, $q=1$ case, large Faraday
rotation acts to randomize a photon's polarization vector between scatterings,
and the limb darkening law approaches that for scattering described by
Rayleigh's phase function (cf. the appendix and section 45 of Chandrasekhar 1960),
i.e. Thomson scattering of unpolarized radiation.  This limb darkening law
turns out to be very close to the pure electron scattering case with polarization
effects.  As shown by the $q=0.8$ curves in figure 4, modest absorption
causes Faraday rotation to have a more substantial effect on the limb darkening,
although it is still small (cf. the $q=0.8$ case in figure 4).
\begin{figure}
\plotone{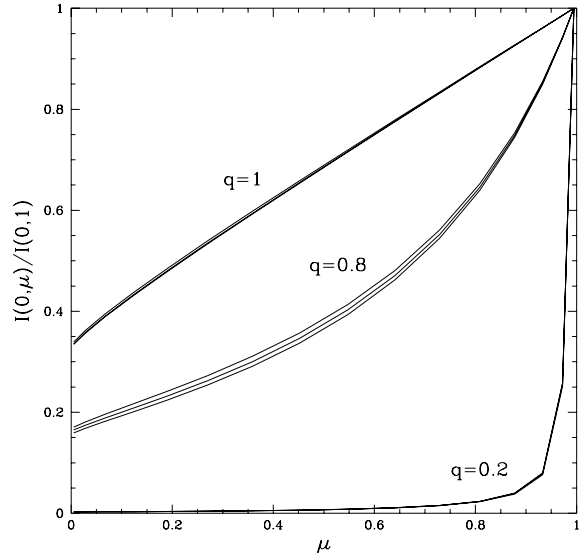}
\caption{Angular distribution of total intensity $I$ emerging from atmospheres
with $q=0.2$, $q=0.8$, and $q=1$; $\delta=0$, 2, and 100; and all radiation
sources at infinite optical depth.
In the $q=0.2$ and $q=1$ cases the limb darkening is virtually independent
of $\delta$,
and the three curves in each case lie nearly on top of each other.  In the
$q=0.8$ case, $\delta$ increases from bottom to top.  All curves were
calculated using our Feautrier code.  Note that for small $q$, the limb
darkening becomes very large, in agreement with equation (\ref{eqlosa}).
}
\end{figure}
This is because the polarization is greater than for the $q=1$ case.
The contribution of $Q$ to the intensity source function is therefore larger,
so as the magnetic field depolarizes, the intensity
source function is modified more than for the $q=1$ case.  For large absorption
opacity, e.g. the $q=0.2$ case shown in figure 4, the effects of Faraday
rotation are very small.

Figure 5 shows the depolarizing effects of Faraday rotation for all values
of $q$.  
\begin{figure}   
\plotone{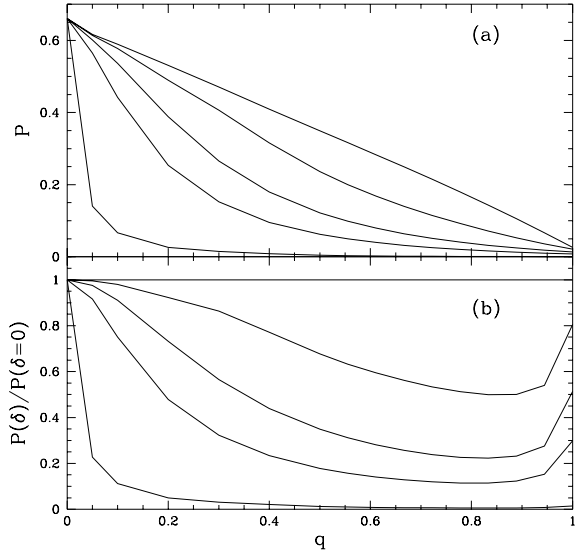}
\caption{Polarization as a function of $q$ along the $\mu=0.452$ line of sight
for various values of Faraday rotation parameter $\delta$.  In (a) we show the
actual polarization, while in (b) we show the ratio of the polarization to
that of the $\delta=0$ (unmagnetized) case.  From top to bottom in both figures,
the curves represent $\delta=0$, 2, 5, 10, and 100.  All the results shown
were calculated using our Feautrier code.
}
\end{figure}
Overall, as $q$ decreases below unity, Faraday rotation is at first
more effective in depolarizing the radiation field than at $q=1$.
The polarized source function has a greater contribution from $Q$ relative
to $I$ than it did in the $q=1$ case.
The limb darkening does not change much with $\delta$,
so as the Faraday rotation is added and the polarization is reduced (except
for $\mu$ near 0), the polarized source function for $\mu=0$ decreases
more rapidly than in the $q=1$ case.
However, as $q$ gets below around 0.4 for the particular viewing angle shown in
figure 5, Faraday rotation starts
to become less effective in depolarizing the radiation field compared to
the $q=1$ case because of the diminishing electron column density down to unit
optical depth.  As $q\rightarrow0$, Faraday rotation has zero effect as we
noted in our analytic solution above.

\subsection{Case 2:  Linear Thermal Source Function}

We now consider a distribution of sources in the atmosphere such that
the thermal source function depends linearly on optical depth,
\begin{equation}
S(\tau)=S(0)(1+\beta\tau),
\end{equation}
where $S(0)$ and $\beta$ are constants.
In the $q\rightarrow1$ limit, this problem reduces again to the
pure electron scattering case considered in paper I.\footnote{This
is true provided the thermal source function remains finite.  By
$q\rightarrow1$,
we mean that $n_e\sigma_T\gg\kappa$ and that the scattering source
function is much larger than the thermal source function.}~
We use the diffusion approximation to apply a lower boundary condition 
at sufficiently high optical depth
$\tau_{\rm max}$ so that the results are independent of $\tau_{\rm max}$.
Apart from the uninteresting
normalization factor, the radiation field emerging from this
atmosphere now depends on three parameters: $\delta$, $q$, and $\beta$.

For atmospheres with very small scattering opacities ($q\rightarrow0$), the
radiative transfer equation may be solved perturbatively.  To lowest order
in $q$, the total intensity from the atmosphere is given by the
Eddington-Barbier relation,
\begin{equation}
I(0,\mu)=S(0)(1+\beta\mu)+O(q),
\end{equation}
and the polarization is given by
\begin{eqnarray}
\label{gnedinpq}
\lefteqn{P(\mu)=q{3(1-\mu^2)\over16(1+\beta\mu)}\left\{\mu\Phi(\mu)+
\beta\left[{1\over4}+\mu^2\Phi(\mu)\right]\right\}}\nonumber\\
& &+O(q^2),
\end{eqnarray}
where
\begin{equation}
\Phi(\mu)\equiv{3\over2}-3\mu+(3\mu^2-1)\ln\left(1+{1\over\mu}\right).
\end{equation}
These results are identical to those obtained by Gnedin \& Silant'ev (1978)
for unmagnetized atmospheres in the $q\rightarrow0$ limit, and this is
because the effects of Faraday rotation are of order $q^2$ in this limit.
Physically, the ratio of scattered to thermal (unpolarized) intensity is
of order $q$, which is why the polarization is also of this
order.  (This is in marked contrast to the previous case we considered where
the thermal sources were all at infinite optical depth.  Here the presence
of a nonzero thermal source function ensures that the polarization vanishes
as $q\rightarrow0$.) The amount by which Faraday rotation can further
depolarize the
radiation is also of order $q$, because this is the factor by which the
electron column density (which does the rotation) down to unit optical depth
is reduced.
Hence we immediately conclude again that Faraday rotation has negligible effect
on the polarization (which is already small) as $q\rightarrow0$.

Figure 6 shows the polarization viewed along $\mu=0.452$ for various values
of $\beta$ and $q$ for an unmagnetized atmosphere.\footnote{The $\beta=\infty$
case corresponds to $S(\tau)\propto\tau$.}~
\begin{figure}
\plotone{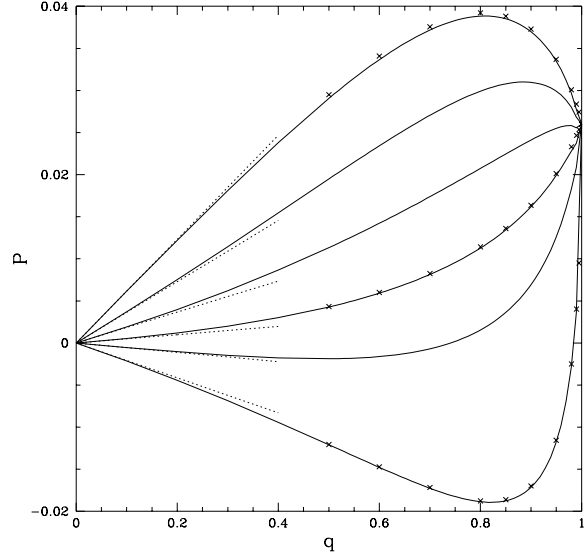}
\caption{Polarization as a function of $q$ along the $\mu=0.452$ line of sight
for various values of source function gradient $\beta$ in an unmagnetized
($\delta=0$) atmosphere.  From bottom to top, the curves represent the results
of our Feautrier calculations for $\beta=0$, 0.5, 1, 2, 5 , and $\infty$.
Points represent the numerical calculations of Loskutov \& Sobolev (1979),
interpolated to $\mu=0.452$.  The dotted lines are the results of the analytic
formula
for small $q$, equation  (\ref{gnedinpq}).  All curves approach the 
Chandrasekhar (1960) value for this viewing angle as $q\rightarrow1$.
}
\end{figure}
Also shown are the numerical
results of Loskutov \& Sobolev (1979), which are again in good agreement with
ours.  Negative values of $P$ in this figure represent cases where the
polarization plane is perpendicular to the plane of the atmosphere.
We call this the ``Nagirner effect'', after the person who first noted that
absorption opacity can produce this (cf. Gnedin \& Silant'ev 1978).  It is
possible that this effect could explain the fact that the observed polarization 
in type I AGN is parallel to the radio axis.  We find that negative
polarization is present for some $\mu$ and $q$ if and only if 
$\beta<(6-8\ln 2)/(8\ln 2-5) = 0.834$.
This is consistent with Gnedin \& Silant'ev's (1978) $q\rightarrow0$ result,
because as $\beta$ drops below the critical value, the polarization first
becomes negative for small $q$.

\begin{figure}
\plotone{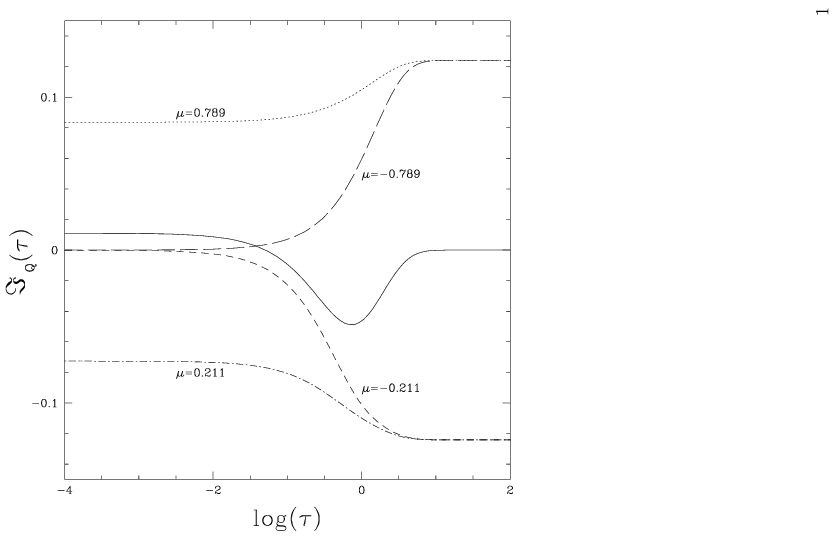}
\caption{Polarized source function vs. optical depth along the $\mu=0.211$
line of sight for $\beta=0$ and $q=0.8$.  The solid curve is the total 
source function, while the
dashed and dotted curves are the contributions from various angles in
the four-stream approximation (negative $\mu$ is downwards). The plot
of $\Im_Q(\tau)$ for the $\mu=0.789$ line of sight is very
similar.} 
\end{figure}
The Nagirner effect therefore arises for sufficiently flat thermal source
functions.  It is useful, however, to examine its origin a little more
closely by considering the depth dependence of the {\it total} source 
function.
If there is no magnetic field, then the outgoing radiation ($0\le\mu\le1$)
can be formally expressed from equation (\ref{eqformfar}) as
\begin{equation}
{\bf I}(0,\mu)={\int_0^\infty\bmath{\Im}(t,\mu)e^{-t/\mu}{dt\over\mu}}.
\end{equation}
Hence if $\Im_Q(\tau,\mu)<0$ over some range of optical depths, then the
outgoing radiation can be negatively polarized for that particular viewing
angle.  The solid curve in figure 7 shows the depth dependence of
$\Im_Q(\tau)$ for $\beta=0$, $q=0.8$, and $\mu=0.211$. 
Around $\tau=1$, $\Im_Q$ is negative, but for low $\tau$
it becomes positive.  The reason for this can be seen by looking at the
contribution to the polarized source function from radiation coming from
different directions in an atmosphere with constant source function.  The
broken curves in figure 7 show the contribution to the source function
from different angles in a four-stream calculation of the radiation field,
i.e. where the radiation is calculated at four angles for the purposes of
computing quadratures.  Near $\tau=0$, there
is no downgoing radiation ($\mu<0$), so the only contribution is
from upgoing radiation.   The limb darkening causes the source function
polarization to
be positive, since the near-vertical radiation (which has net positive
polarization when scattered) is stronger than the near-horizontal radiation
(which has net negative polarization when scattered).  Near $\tau=1$,
the limb darkening is weaker.  Here, there begins to be a significant
contribution to the source function from downward radiation which is
produced in the layers above $\tau=1$.  The near-horizontal radiation is
stronger than the near-vertical since there is more atmosphere emitting
from smaller $|\mu|$.  This leads to a net negative
polarization at $\tau=1$.  Figures 8 and 9 illustrate this effect. (Note that
the polarization is low, so most of the contribution to $\Im_Q$ comes
from $I$.  Thus, the corresponding difference in the contribution to $\Im_Q$
from different angles is due mostly to the difference in the intensities
from different angles.)
When the thermal source function has a steep vertical gradient, the limb
darkening is stronger.  The contribution from the downward going radiation
is less than that of
the upward radiation, and the polarization source function is always
positive.  This is why flat thermal source function gradients are required
for the Nagirner effect to be present.
\begin{figure}
\plotone{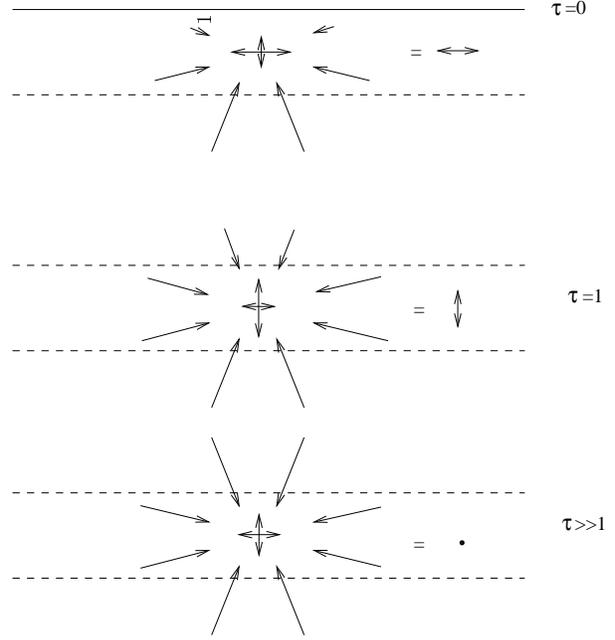}
\caption{Cartoon showing the reason why the polarized source function switches 
from positive to negative to zero with increasing $\tau$. The length
of the radial arrows represents the strength of radiation coming
from different angles.  The arrows in the centre represent the
strength of the negative and positive contributions to the polarized
scattering source function (negative is vertical, positive is horizontal).
The sum of the polarizations is indicated on the right, along with the
optical depth of each layer.
}
\end{figure}
\begin{figure}
\plotone{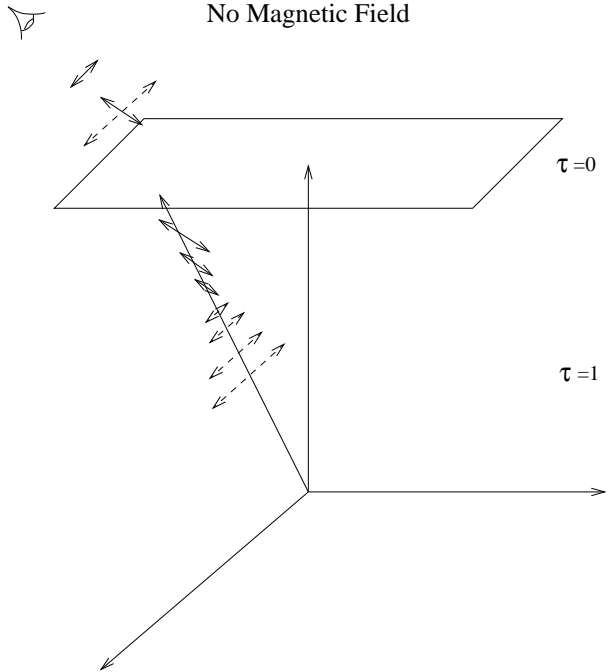}
\caption{At wavelengths where the Nagirner effect is present, the polarization
vector lies in the plane of the line of sight and the vertical
at $\tau=1$, while it is perpendicular near $\tau=0$.  Thus, the polarization
ends up being negative at some viewing angles.  However, when a magnetic field
is added, Faraday rotation causes the polarization near $\tau=1$ to be rotated
and depolarized, while the radiation near $\tau=0$ is rotated very little, so
the polarization ends up being nearly perpendicular.}
\end{figure}

Figure 10 shows the polarization as a function of viewing angle for
$\beta=1$ and (a) $q=0.8$ and (b) $q=0.2$.  This figure
should be compared with figure 2 above.  Faraday rotation has a much smaller
effect on the polarization at $\mu=0$ when a nonzero thermal source function
is present.  This is because the polarization is low and limb darkening is
more important in determining the polarized source function.
In addition, the presence of the thermal
source function implies that the overall polarization vanishes as
$q\rightarrow0$.  Faraday rotation has even less effect on the polarization,
which is already small, as $q\rightarrow0$.  For example, in the $q=0.2$ case
shown in figure 10(b), the $\delta=0$, 2, and 5 curves overlap because the
effects of Faraday rotation are reduced by an additional factor of $q$ as
discussed above.
\begin{figure}
\plotone{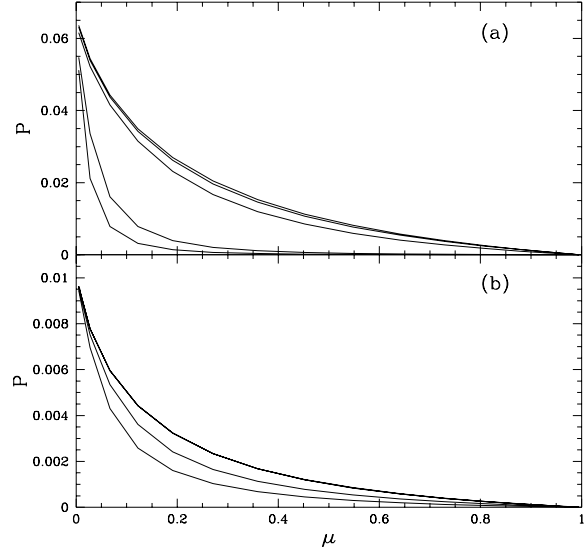}
\caption{ Polarization as a function of viewing angle for (a) $q=0.8$ and
(b) $q=0.2$
atmospheres with various values of $\delta$ and a linear source function
with $\beta=1$.  From top to bottom,
the curves represent the results of our Feautrier calculations for $\delta=0$,
2, 5, 50, and 100.
}
\end{figure}

Figure 11 shows the polarization in the $\beta=1$ and $\infty$ cases as a
function of $q$ for various $\delta$.  Notice again that as $q$ becomes small,
the effect of the Faraday rotation decreases, and all curves approach the
$\delta=0$ case.  Note however that this approach is much smoother than in
the previous case where all the sources were at infinite optical depth
(cf. figure 5).  For moderate absorption opacity ($q$ slightly less than unity),
the depolarizing effects of Faraday rotation are again enhanced over the pure
scattering problem for the $\beta=\infty$ case shown.  This is the same effect
as in section 3.1, and again arises because the absorption opacity on its own
increases the polarization.  If the thermal source function gradient is not
so steep so that modest absorption opacity decreases the polarization, then
the effects of Faraday rotation are reduced as $q$ drops below unity (cf. the
$\beta=1$ case in figure 11).
\begin{figure}
\plotone{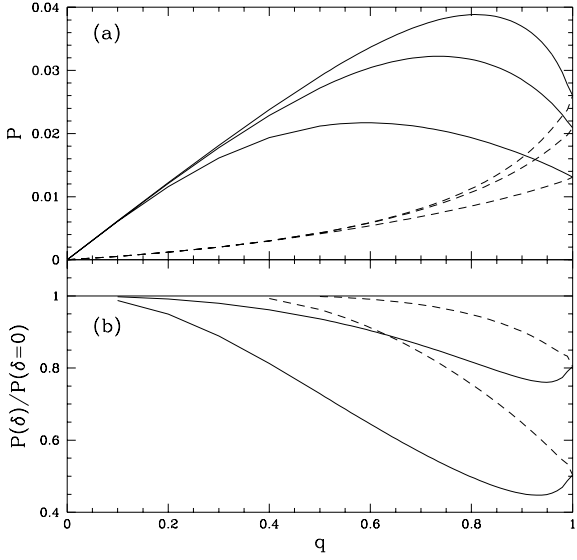}
\caption{The effect of Faraday rotation
on the $\beta$= 1 (dashed) and $\infty$ (solid) cases for the $\mu=0.452$ line
of sight.  In (a) we show the actual polarization, while in (b) we show
the ratio of the polarization to that of the $\delta=0$ (unmagnetized) case.
From top to bottom in both figures, the curves represent the results of our
Feautrier calculations for $\delta=0$, 2, and 5.}
\end{figure}

So far we have discussed Faraday rotation as an agent for depolarizing the
radiation field.  However, when $\beta$ is small and the Nagirner effect is
present, it is possible
for the magnetic field to {\it increase} the polarization.  Figure 12 shows
the normalized Stokes parameters
for $q=0.9$ and $\beta=0.25$ for various $\delta$, and $Q/I$ with the 
$\delta=0$ case subtracted.  
\begin{figure}
\plotone{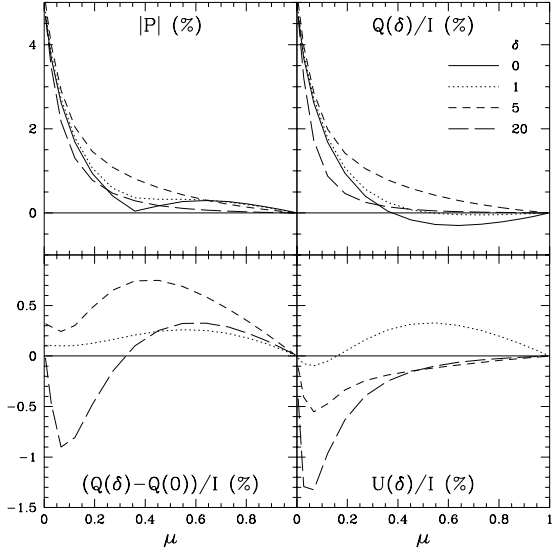}
\caption{The top panel shows $P$ and $Q/I$, and the bottom panel
$(Q(\delta)-Q(0))/I$ and $U/I$ as 
$\delta$ increases in an atmosphere with $q=0.9$ and $\beta=0.25$.  
Note that at large $\delta$
and small $\mu$, $Q/I$ decreases relative to the zero magnetic field case,
which is because the Faraday rotation is significant for small optical
depths when $\delta$ is large enough.}
\end{figure}
With no magnetic field, the $\Im_Q$'s from
different depths partially cancel, making the polarization negative for
$\mu>0.38$, but positive for smaller $\mu$.

Since $\Im_U=0$, the
only contribution to the polarization is from $\Im_Q$, which for $q=0.9$ and
$\beta=0.25$ is similar
to the source function plotted in figure 7.  Because the polarization is low,
$\Im_Q$ does not change much with $\delta$ here.  For this case, $\Im_Q < 0$
for $\tau > 0.25$ at all $\mu$.  Thus, the positive polarized source
function comes 
from a small range of electron scattering depths, so the Faraday
depolarization is insignificant for $\delta\lesssim4$.  However, the 
negative polarized source function comes from a larger range of optical depths, 
causing Faraday depolarization for $\delta\simeq1$, reducing the magnitude of
the negative polarization so that the outgoing polarized flux is increased.
\footnote{Strictly speaking, ``positive'' and ``negative''
polarization define the orientation of the plane of
polarization only when there is no magnetic field present, because there
are only two possible orientations.  The presence of the magnetic field
breaks the azimuthal symmetry and allows the plane of polarization to
be at an arbitrary angle with respect to the vertical/line of sight
plane.  We use these terms here in reference to the unmagnetized case to show
how Faraday rotation acts on the different orientations of the polarization
at different depths in the atmosphere.}~  
For very large $\delta$, even the
radiation from small $\tau$ will be Faraday rotated significantly, so
the polarization will still be approximately horizontal, but will eventually
start to decrease in magnitude again, as seen in figure 12.
The effect of increasing the polarization with increased magnetic field
in a semi-infinite atmosphere is only present when there is absorption
opacity present and a shallow source function gradient. 

\section{Polarization from realistic atmospheres in an accretion disk}

In a previous paper (Blaes \& Agol 1996), we calculated the structure of
local, static, plane-parallel atmospheres using the complete linearization
technique, neglecting the effects of any magnetic field.\footnote{In particular,
we ignore the contribution of magnetic field pressure on hydrostatic equilibrium,
which may be quite important for equipartion fields.}~ Here we
use some of these atmospheres to calculate the radiative transfer in
the presence of a constant vertical magnetic field in the atmosphere according
to equation (\ref{eqtransfer}).
Our atmosphere solutions include hydrogen bound-free and free-free
opacities as well as electron scattering opacity.  Non-LTE effects in the
$n=1$ and 2 levels of hydrogen are included.  The results with magnetic fields
of different strengths in atmospheres with ($T_{eff}$, $g$) of
($2\times10^4$~K, 190~cm~s$^{-2}$), ($4.5\times10^4$~K,
$4\times10^3$~cm~s$^{-2}$), and ($10^5$~K, $9.5\times10^4$~cm~s$^{-2}$) 
are plotted in figures 13-15, respectively.  We also show the variation of $q$
at $\tau=1$ with wavelength in figure 16 for these three atmospheres.
In all cases the magnetic field does not
significantly affect the total intensity spectrum $I_\lambda$.
\begin{figure}
\plotone{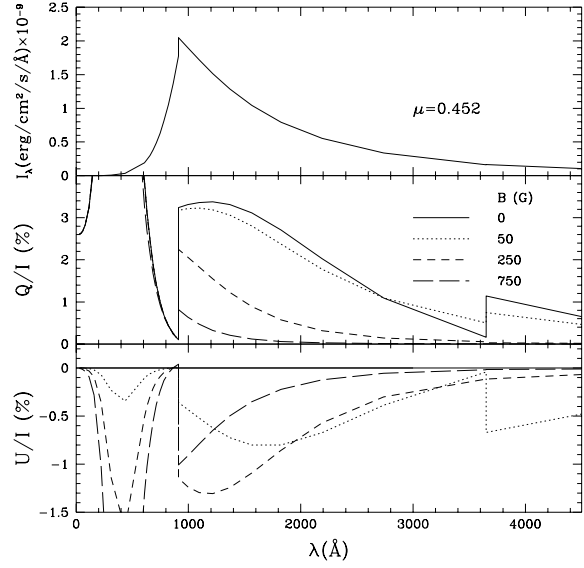}
\caption{Stokes parameters in an atmosphere including Faraday rotation for
$T_{eff}=20,000$~K and $g=190$~cm~s$^{-2}$ including non-LTE effects for
two hydrogen levels.  The curves depicted are for a $\mu=0.452$ line of
sight, with various magnetic field strengths.  The top panel shows the
outgoing total intensity spectrum $I_\lambda=\nu/\lambda I_\nu$.
The bottom two panels show the percent polarization for the $Q/I$ and
$U/I$ Stokes parameters. $Q/I$ peaks at 10 per cent and
$U/I$ peaks at -4 per cent at about 500${\rm\AA}$ for $B=750$~G.} 
\end{figure} 
\begin{figure}
\plotone{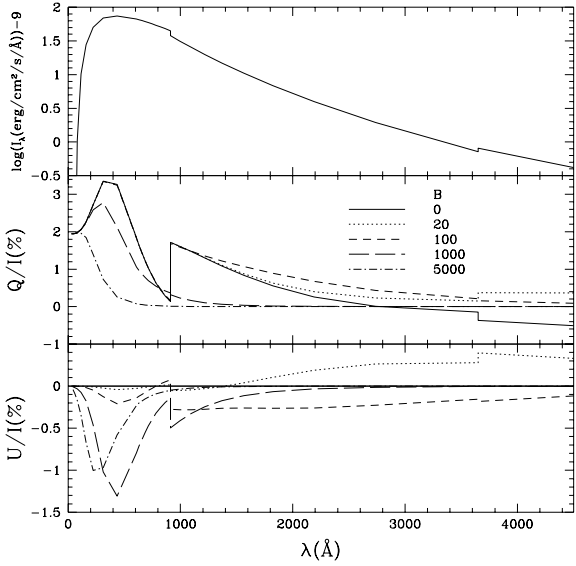}
\caption{Polarization in an atmosphere including Faraday rotation for
$T_{eff}=45,000$K, $g=4000$~cm~s$^{-2}$, and $\mu=0.55$ and various 
magnetic field strengths.}
\end{figure} 

As shown in figure 16, $q$ is very small just blueward of the Lyman edge,
so the Faraday rotation does not affect the polarization very much in this
region of the spectrum unless $\delta$ is very large.  This is illustrated
in figures 13-15.  In the case shown in figure 13, the source function is
shallow just blueward of the Balmer edge. Hence $\Im_Q<0$
near $\tau=1$ at these wavelengths.  Here the increasing magnetic
field causes an {\it increase} in polarization, as described in section 3.
Figure 13 also shows a dramatic reduction in the difference in polarization
across the Lyman edge. This is due to the fact that
redward of the edge, $q_\nu$ is large, so the Faraday depolarization is
strong.  Blueward of the edge, however, $q_\nu$ is small, so the Faraday
depolarization is weak.  Thus, as the magnetic field increases,
the polarization decreases faster redward than blueward of the edge.
In some cases the polarization blueward of the edge is larger than
redward of the edge, as shown in figure 15 for $B=2500$~G.

The Balmer edge in figure 14 has a negative polarization,
which becomes more negative redward of the edge for $B=0$. 
However, as the magnetic field increases, the polarization first becomes
positive, and increases from blue to red across the edge. For larger B, 
the polarization is reduced, decreasing from blue to red.  This behavior 
can be understood in terms of the physics discussed in section 3.  The 
thermal source functions are very flat on both sides of the Balmer edge, 
but $q$ is larger redward of the edge ($q\simeq0.8$ at $\tau=1$) than
blueward of the edge ($q\simeq0.57$ at $\tau=1$).  Hence the Nagirner effect
is stronger (cf. figure 6).  Since Faraday depolarization is stronger for
larger $q$, the polarization increases to the red across the edge when there
is a 20~G magnetic field ($\delta=2.1$ at $3648{\rm\AA}$).  However, when
$\delta$ is larger, the depolarization is more rapid for larger $q$, so
then the polarization decreases to the red across the edge when $B=100$~G,
or $\delta=10.5$ at $3648{\rm\AA}$.

If the magnetic field were randomly oriented, 
the polarization feature may also be decreased at larger
inclination angles for a large magnetic field.

\begin{figure}
\plotone{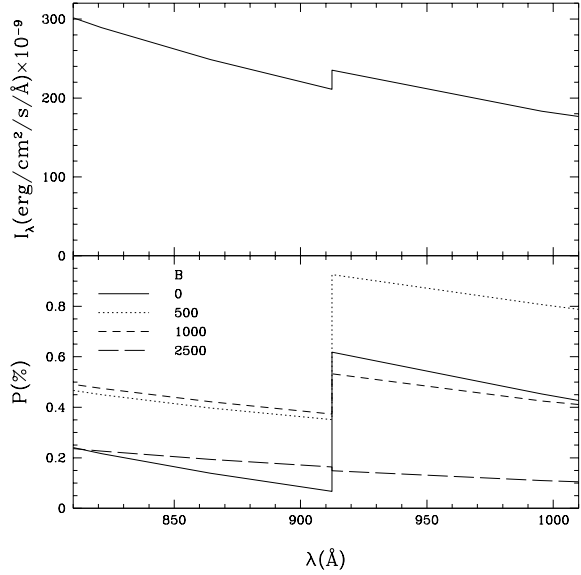}
\caption{Polarization near the Lyman edge in an atmosphere including 
Faraday rotation for $T_{eff}=100,000$~K, $g=95,000$~cm~s$^{-2}$, and
$\mu=0.55$ and various magnetic field strengths.}
\end{figure} 
\begin{figure}
\plotone{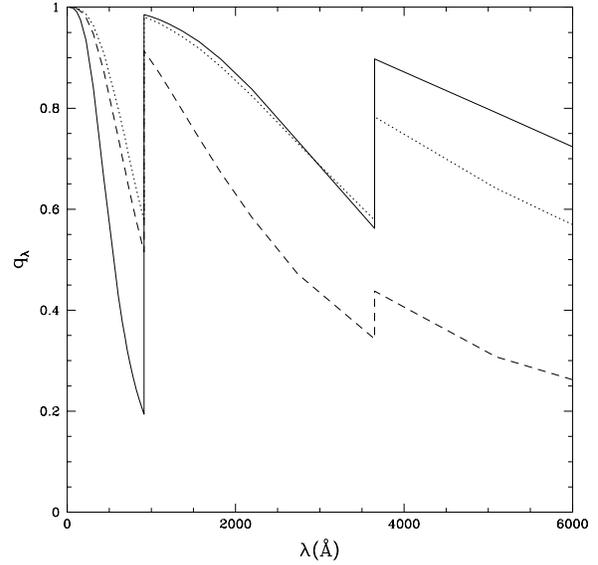}
\caption{The ratio of scattering opacity to total opacity at $\tau=1$
for the three atmospheres displayed in figures 13-15.  The solid,
dotted, and dashed curves represent $q_\lambda$ for $T_{eff}$= 20,000K,
45,000K, and 100,000K atmospheres respectively.}
\end{figure}

The results of this section are meant to give a flavor of the complexities
that may result in the polarized spectrum emerging from a realistic atmosphere.
The full accretion disk spectrum requires an integration over many such
atmospheres in different physical conditions representing the disk at different
radii.

\section{Conclusions}

Using a combination of numerical calculations and analytic arguments, we
have extended our previous study of Faraday rotation in accretion disk
atmospheres to include the interaction with absorption opacity.  Along
the way we have clarified the role that each of these two effects play
separately.  Faraday rotation in a pure electron scattering atmosphere acts
to depolarize the radiation field by rotating the polarization vectors
of photons scattered from different depths.

Absorption opacity in an unmagnetized atmosphere can increase or decrease the
polarization depending on the behavior of the thermal source function.  If the
thermal source function is zero except at great depth, absorption opacity alone
always increases the polarization.  In the more usual case of nonvanishing
thermal source function, the effect of absorption opacity depends on the source
function gradient.  If the thermal source function increases steeply with
depth, absorption opacity can increase the polarization over the pure electron
scattering case.  On the other hand if the source function increases slowly
with depth, or even decreases, then absorption opacity can flip the plane
of polarization to be in the vertical/line of sight plane.  While these
results were known from previous work (Gnedin \& Silant'ev 1978, Loskutov
\& Sobolev 1979), we have presented a novel physical interpretation in terms
of the behavior of the total source function $\bmath\Im$.  Quite generally
this source function locally produces polarization which is parallel to the 
atmosphere plane at low optical depths.  At optical depths around unity,
however, the polarization can be parallel or perpendicular to the atmosphere
plane depending on the thermal source function gradient.

When Faraday rotation is combined with absorption opacity in a scattering
atmosphere, a number of effects can occur.  First, if absorption dominates
scattering, then the electron column density along a photon mean free path
is small and Faraday rotation only has a small effect on the polarization
of the radiation field.  On the other hand, if modest absorption opacity
alone increases the polarization, then the depolarizing effects of Faraday
rotation are enhanced compared to the pure scattering
case, at least for a vertical magnetic field.
On the whole Faraday rotation generally acts to depolarize the radiation field,
as in the pure scattering case.  However, when the Nagirner effect is
present at certain photon wavelengths, then Faraday rotation can actually
increase the emerging polarization by depolarizing the deeper radiation
field which has a perpendicular orientation to the radiation which is
scattered from shallower depths.

While we have shown that these effects can all occur in simple toy model
atmospheres, we have also demonstrated that they are present in more
realistic atmospheres.  We have also shown how the polarized radiative
transfer can be computed in a straightforward manner by a simple extension
of the Feautrier method to incorporate a vertical magnetic field.
This numerical method can be used to integrate the radiation
field produced at different annuli in the accretion disk to calculate the
total observed radiation field.  As we noted in paper I, the largest
uncertainty in applying such calculations to the observed data is the
variation of magnetic field strength with disk radius, and its covering
factor on the disk photosphere.  (It might be possible to get a handle
on the latter by combining ultraviolet and X-ray observations to determine
the ``patchiness'' of the corona; cf. Haardt,
Maraschi, \& Ghisellini 1994.  This assumes that such patches, if real, are
magnetized active regions similar to those on the sun.)    The field topology
will of course also not be vertical, but following
our Monte Carlo work of paper I, we expect magnetic fields of random
orientation to have qualitatively (and perhaps quantitatively to some extent)
similar effects.

\section*{Acknowledgments}

We are grateful to Ari Laor for emphasizing to us the role that absorption
opacity might play in modifying the effects of Faraday rotation, which
helped initiate the research reported here.
This work was supported by NSF grant AST~95-29230.

\appendix
\section{}
We reproduce here Silant'ev's (1979) analytic calculation of the polarization
of radiation emerging from an optically thick, pure electron scattering
atmosphere in the $\delta\rightarrow\infty$ limit.

Silant'ev's calculation uses a radiation density matrix formalism, which is
applicable to magnetic fields with arbitrary orientation.  For a vertical
magnetic field, however, it is somewhat simpler to use the radiative
transfer equation as we have formulated it in section 2.  We therefore
proceed to show how Silant'ev's result can be derived within this formalism.

Define a complex intensity column vector
\begin{equation}
{\bf I'}=\pmatrix{I/2^{1/2} \cr (-Q+iU)/2 \cr (-Q-iU)/2}.
\end{equation}
Then, for $q=1$, the equation of transfer (\ref{eqtransfer}) may be written
\begin{equation}
\mu{\partial{\bf I'}\over\partial\tau}={\bf M}{\bf I'}
-{1\over2}\int_{-1}^1d\mu'{\bf P'}(\mu,\mu'){\bf I'}(\mu'),
\end{equation}
where
\begin{equation}
\label{mmatrix}
{\bf M}=\pmatrix{1 & 0 & 0 \cr 0 & 1\mp i\delta\mu & 0 \cr 0 & 0 &
1\pm i\delta\mu},
\end{equation}
\begin{equation}
\label{eqpprime}
{P'}_{ij}(\mu,\mu')=a_i(\mu)a_j(\mu')+ \delta_{i1}\delta_{j1},
\end{equation}
and\
\begin{equation}
{\bf a}(\mu)=\left(-P_2(\mu)/2^{1/2},3(1-\mu^2)/4,3(1-\mu^2)/4\right).
\end{equation}
As in section 2.2, the upper and lower signs in equation (\ref{mmatrix}) refer
to upward and downward magnetic field directions, respectively.  We have again
suppressed the frequency subscript because there is no
frequency redistribution in this problem.
 
Using a principle of invariance for vectors (cf. chapter 4 in Chandrasekhar 
1960), the radiation field
emerging from the atmosphere ($0\le\mu\le1$) can be expressed in terms of a
scattering matrix {\bf S}:
\begin{eqnarray}
\label{outflux}
\lefteqn{{\bf I'}(0,\mu)=\pmatrix{1& 0 & 0 \cr 0 & {1\over 1\mp i\delta\mu}
& 0 \cr
0 & 0 & {1\over 1\pm i\delta\mu}}\left[{1\over2}\int_0^1{\bf P'}(\mu,\mu')
{\bf I'}(0,\mu')d\mu'\right.}
\nonumber\\
\lefteqn{+\left.{1\over4}\int_0^1\int_0^1{\bf S}(\mu,\mu')
{\bf P'}(\mu',\mu''){\bf I'}(0,\mu''){d\mu'\over\mu'}d\mu''\right].}
\end{eqnarray}
The scattering matrix satisfies the following integral equation:
\begin{eqnarray}
\label{scateqn}
\left({1\over\mu}+{1\over\mu_0}\right){\bf S}(\mu,\mu_0)\mp
i\delta{\bf D}{\bf S}(\mu,\mu_0)\pm i\delta{\bf S}(\mu,\mu_0){\bf D}
\nonumber \\
={\bf P'}(\mu,\mu_0)+{1\over2}\int_0^1{\bf P'}(\mu,\mu'){\bf S}(\mu',\mu_0)
{d\mu'\over\mu'} +
\nonumber \\
{1\over2}\int_0^1{\bf S}(\mu,\mu'){\bf P'}(\mu',\mu_0)
{d\mu'\over\mu'}+
\nonumber \\
{1\over4}\int_0^1\int_0^1{\bf S}(\mu,\mu'){\bf P'}(\mu',\mu'')
{\bf S}(\mu'',\mu_0){d\mu'\over\mu'}{d\mu''\over\mu''},
\end{eqnarray}
where
\begin{equation}
{\bf D}=\pmatrix{0 & 0 & 0 \cr 0 & 1 & 0 \cr 0 & 0 & -1}.
\end{equation}
Note that the symmetry of equations (\ref{eqpprime}) and (\ref{scateqn})
implies that ${\bf S}(\mu,\mu_0)={\bf S}(\mu_0,\mu)$.

Define 
\begin{equation}
\bgamma(\mu,\mu_0)\equiv \mu+\mu_0 
\pmatrix{1 & 1\pm i\delta\zeta &
1\mp i\delta\zeta \cr 1\mp i\delta\zeta & 1 &
1\mp 2i\delta\zeta \cr 1\pm i\delta\zeta &
1 \pm 2i\delta\zeta & 1},
\end{equation}
(where $\zeta=\mu\mu_0/(\mu+\mu_0)$)
and a matrix ${\bf T}(\mu,\mu_0)$ by letting
\begin{equation}
S_{ij}(\mu,\mu_0)\equiv{\mu\mu_0\over\gamma_{ij}(\mu,\mu_0)}T_{ij}(\mu,\mu_0),
\end{equation}
where no summation over the indices $i$ and $j$ is implied.  In the limit of
large $\delta$, $1/\gamma_{ij}(\mu,\mu_0)=\delta_{ij}/(\mu+\mu_0)$,
provided both $\mu$ and $\mu_0$ are nonzero.  If we make the ansatz that all
elements of the matrix ${\bf T}$ remain finite for all values of $\mu$
and $\mu_0$, then in this limit equation (\ref{scateqn}) simplifies to
\begin{equation}
T_{ij}(\mu,\mu_0)=H_i(\mu)H_j(\mu_0)+K_i(\mu)K_j(\mu_0).
\end{equation}
Here
\begin{equation}
H_i(\mu)=a_i(\mu)+{\mu\over2}\int_0^1{T_{ii}(\mu,\mu')a_i(\mu')
\over \mu+\mu'} d\mu',
\end{equation}
\begin{equation}
K_i(\mu)=\delta_{i1}\left[1+{\mu\over 2}\int_0^1{T_{11}(\mu,\mu')
\over \mu+\mu'}d\mu'\right],
\end{equation}
and we are considering both $\mu$ and $\mu_0$ to be nonzero.
Since $a_2(\mu)=a_3(\mu)$, the equations for $T_{22}$ and $T_{33}$ are
identical.  Let $H_2(\mu)=H_3(\mu)\equiv{3\over4}(1-\mu^2)\phi(\mu)$.  Then
$\phi(\mu)$ satisfies the nonlinear equation
\begin{equation}
\phi(\mu)=1+{9\mu\phi(\mu)\over32}\int_0^1{\phi(\mu')(1-\mu'^2)^2
\over \mu+\mu'}d\mu'.
\end{equation}
The equation for $T_{11}$ may be written
\begin{equation}
\label{eqt11}
T_{11}(\mu,\mu_0)={3\over8}\left[{1\over3}\psi(\mu)\psi(\mu_0)+{8\over3}
\tilde\phi(\mu)\tilde\phi(\mu_0)\right],
\end{equation}
where
\begin{equation}
\psi(\mu)\equiv3-\mu^2+{\mu\over2}\int_0^1{T_{11}(\mu,\mu')(3-\mu^2)\over
\mu+\mu'}d\mu'
\end{equation}
and
\begin{equation}
\tilde\phi(\mu)\equiv\mu^2+{\mu\over2}\int_0^1{T_{11}(\mu,\mu'){\mu'}^2\over
\mu+\mu'}d\mu'.
\end{equation}
Apart from multiplicative factors, equation (\ref{eqt11}) is identical to
the equation for the first term of the scattering function $S^{(0)}(\mu,\mu_0)$
in the problem of diffuse reflection from a scattering medium described by
Rayleigh's phase function (cf. equation 6 in section 44 of Chandrasekhar
1960).

The scattering matrix elements in the limit of large $\delta$ are therefore
given by
\begin{equation} \nonumber
\lefteqn{S_{11}(\mu,\mu_0)={3\over8}{\mu\mu_0\over\mu+\mu_0}
H(\mu)H(\mu_0)[3-c(\mu+\mu_0)+\mu\mu_0]}
\end{equation}
\begin{eqnarray}
S_{22}(\mu,\mu_0)&=&S_{33}(\mu,\mu_0)\nonumber\\
&=&{9\over 16}{\mu\mu_0\over\mu+\mu_0}
(1-\mu^2)(1-\mu_0^2)\phi(\mu)\phi(\mu_0)
\end{eqnarray}
where $H(\mu)$ satisfies the integral equation,
\begin{equation}
H(\mu)=1+{3\over16}\mu H(\mu)\int_0^1{(3-\mu'^2)\over \mu+\mu'}H(\mu')d\mu',
\end{equation}
and
\begin{equation}
c\equiv{\int_0^1H(\mu)\mu^2d\mu\over\int_0^1H(\mu)\mu d\mu}.
\end{equation}
The off-diagonal components of ${\bf S}(\mu,\mu_0)$ are negligible
$(\sim\delta^{-1})$.  Although these expressions
were derived assuming both $\mu$ and $\mu_0$ are nonzero, they remain valid
even if this is not the case.  This is because they then vanish provided
${\bf T}(\mu,\mu_0)$ is finite, which can easily be shown to be true.

Given this solution for ${\bf S}(\mu,\mu_0)$, we can calculate the
outgoing radiation field from equation (\ref{outflux}).  We immediately
conclude from the matrix multiplier in this equation that $Q$ and $U$ vanish
as $\delta\rightarrow\infty$ for nonzero $\mu$.  The intensity is given by
\begin{eqnarray}
I(0,\mu)={1\over2}\int_0^1{P'}_{11}(\mu,\mu')I(0,\mu')d\mu' \nonumber \\
+{1\over4}\int_0^1\int_0^1S_{11}(\mu,\mu')
{P'}_{11}(\mu',\mu'')I(0,\mu''){d\mu'\over\mu'}d\mu'',
\end{eqnarray}
which has solution
\begin{equation}
I(0,\mu)={FH(\mu)\over2\pi\int_0^1\mu'H(\mu')d\mu'},
\end{equation}
where $F$ is the flux emerging from the atmosphere.  This is the same
limb darkening law as that of radiation emerging from a scattering medium
described by Rayleigh's phase function (cf. section 45 of Chandrasekhar 1960),
i.e. an electron scattering medium in which the radiation is everywhere
treated as unpolarized.  This makes physical sense in the limit of infinite
Faraday rotation.

Since the $Q'$ and $U'$ Stokes parameters are of minimum order $\delta^{-1}$
(for $\mu$ not equal to 0), we can ignore the contribution of $Q'$ and $U'$
in the integrals on the right hand side of equation (\ref{outflux}).  This 
then gives:
\begin{eqnarray}
Q'={1\over 1-i\delta\mu}\left[{1\over2}\int_0^1 P_{21}'(\mu,\mu')
I'(0,\mu')d\mu'\right.\nonumber\\
\left.+{1\over4}\int_0^1\int_0^1 S_{22}(\mu,\mu')P_{21}'(\mu',
\mu'')I'(0,\mu''){d\mu'\over\mu'}d\mu''\right],
\end{eqnarray}
and
\begin{equation}
U'=Q'^*.
\end{equation}
Performing these integrals, using equation (42), and solving for $Q$ and $U$, 
we get:
\begin{equation}
Q={3(1-\mu^2)\phi(\mu)\over 8(1+\delta^2\mu^2)}\int_0^1P_2(\mu')
I(0,\mu')d\mu',
\end{equation}
and
\begin{equation}
U={3\delta\mu(1-\mu^2)\phi(\mu)\over 8(1+\delta^2\mu^2)}
\int_0^1P_2(\mu')I(0,\mu')d\mu',
\end{equation}
which are the same as the results obtained by Silant'ev (1979).
The polarization at $\mu=0$ is independent of $\delta$, even for 
$\delta=\infty$, when the polarization is zero everywhere except 
$\mu=0$ where it is equal to 9.137 per cent.
We completely retract our claim in paper I that Silant'ev's treatment breaks
down near $\mu=0$.  This was based on inaccuracies in our Monte Carlo
results at that time.

\label{lastpage}

\end{document}